\documentclass[]{piparticle-final}
\usepackage{graphicx}
\usepackage{amsmath}

\begin{document}

\volume{1}               
\articlenumber{010005}   
\journalyear{2009}       
\editor{J. A. Bertolotto}   
\received{9 July 2009}     
\accepted{6 November 2009}   
\runningauthor{P. Longone \itshape{et al.}}  
\doi{010005}         

\title{The effect of the lateral interactions on the critical behavior
of long straight rigid rods on two-dimensional lattices}

\author{P. Longone,\cite{inst1}
        D. H. Linares,\cite{inst1}
        A. J. Ramirez-Pastor\cite{inst1}\thanks{E-mail: antorami@unsl.edu.ar}
}

\pipabstract{ Using Monte Carlo simulations and finite-size
scaling analysis, the critical behavior of attractive rigid rods
of length $k$ ($k$-mers) on square lattices has been studied. An
ordered state, with the majority of $k$-mers being horizontally or
vertically aligned, was found. This ordered phase is separated
from the disordered state by a continuous transition occurring at
a critical density $\theta_c$, which increases linearly with the
magnitude of the lateral interactions.} \maketitle

\blfootnote{
\begin{theaffiliation}{99}
   \institution{inst1} Departamento de F\'{\i}sica, Instituto de F\'{\i}sica
Aplicada, Universidad Nacional de San Luis-CONICET, Chacabuco 917,
D5700BWS San Luis, Argentina.
\end{theaffiliation}
}

\section{Introduction}

The study of systems of hard non-spherical colloidal particles has, for many years,
been attracting a great deal of interest and the
activity in this field is still
growing~[1--14]. 
An early seminal contribution to this subject was made by
Onsager~\cite{ONSAGER} with his paper on the isotropic-nematic
(I-N) phase transition in liquid crystals. The Onsager's theory
predicted that very long and thin rods interacting with only
excluded volume interaction can lead to long-range orientational
(nematic) order. Thus, at low densities, the molecules are
typically far from each other and the resulting state is an
isotropic gas. However, at large densities, it is more favorable
for the molecules to align spontaneously (there are many more ways
of placing nearly aligned rods than randomly oriented ones), and a
nematic phase is present at equilibrium.

Interestingly, a number of papers have appeared recently, in which
the I-N transition was studied in two
dimensions~[10--14]. 
 In
Ref.~\cite{GHOSH}, the authors gathered strong numerical evidence to suggest
that a system of square geometry, with two allowed orientations,
shows nematic order at intermediate densities for $k \geq 7$ and
provided a qualitative description of a second phase transition
(from a nematic order to a non-nematic state) occurring at a
density close to $1$. However, the authors were not able to
determine the critical quantities (critical point and critical
exponents) characterizing the I-N phase transition occurring in
the system. This problem was resolved in
Refs.~\cite{EPL1,PHYSA19}, where an accurate determination of the
critical exponents, along with the behavior of Binder cumulants,
showed that the transition from the low-density disordered phase
to the intermediate-density ordered phase belongs to the 2D Ising
universality class for square lattices and the three-state Potts
universality class for honeycomb and triangular lattices. Later,
the I-N phase transition was analyzed by combining Monte Carlo
(MC) simulations and theoretical analysis~\cite{JSTAT1,JCP7}. The
study in Refs.~\cite{JSTAT1,JCP7} allowed (1) to obtain $\theta_c$
as a function of $k$ for square, triangular and honeycomb
lattices, being $\theta_c(k) \propto k^{-1}$ (this dependence was
already noted in Ref.~\cite{GHOSH}); and (2) to determine the
minimum value of $k$ ($k_{min}$), which allows the formation of a
nematic phase on triangular ($k_{min}=7$) and honeycomb
($k_{min}=11$) lattices.

In a recent paper, Fischer and Vink \cite{FISCHER} indicated that
the transition studied in Refs.
[10--14] 
corresponds to a liquid-gas
transition, rather than I-N. This interpretation is consistent
with the 2D-Ising critical behavior observed for monodisperse
rigid rods on square lattices \cite{EPL1}. This point will be
discussed in more detail in Sec. III.

In contrast to the systems studied in
Refs.~[10--14]
, many rod-like
biological polymers are formed by monomers reversibly
self-assembling into chains of arbitrary length. Consequently, these
systems exhibit a broad equilibrium distribution of filament
lengths. A model of self-assembled rigid rods has been recently
considered by Tavares et al.~\cite{TAVARES}. The authors focused
on a system composed of monomers with two attractive (sticky)
poles that polymerize reversibly into polydisperse chains and, at
the same time, undergo a continuous I-N phase transition. The
obtained results revealed that nematic ordering enhances bonding.
In addition, the average rod length was described quantitatively
in both phases, while the location of the ordering transition,
which was found to be continuous, was predicted semiquantitatively
by the theory.

Beyond the differences between lattice geometry and the characteristics of
the rods (self-assembled or not), one
fundamental feature is preserved in all the studies mentioned
above. This is the assumption that only excluded volume
interactions between the rods are considered (except in
Ref.~\cite{TAVARES}, where monomers with two attractive bonding
sites polymerize into polydisperse rods). Moreover, one often
encounters phrases in the literature, such as ``This theory
[Onsager's theory] shows that repulsive interactions [excluding
volume interactions] alone can lead to long-range orientational
nematic order, disproving the notion that attractive interactions
are a prerequisite"~\cite{WENSINK}, which could be ambiguous with
respect to the role that attractive lateral interactions between
the rods should play in reinforcing (or not) the nematic order.

In this context, it is of interest and of value to inquire how the
existence of lateral interactions between the rods influences the
phase transition occurring in the system. The objective of this
paper is to provide a thorough analysis in this direction. For
this purpose, an exhaustive study of the phase transition
occurring in a system of attractive rigid rods deposited on square
lattices was performed. The results revealed that $(i)$ the
orientational order survives in the presence of attractive lateral
interactions; $(ii)$ the critical density shifts to higher values
as the magnitude of the lateral interactions is increased; and
$(iii)$ the continuous transition becomes first order for
interaction strength $w > w_c$ (in absolute values).

The outline of the paper is as follows. In Sec. II we describe the
lattice-gas model and the simulation scheme. In Sec. III we
present the MC results. Finally, the general conclusions are given
in Sec. IV.

\section{Lattice-gas model and Monte Carlo simulation scheme}

\begin{figure}
\begin{center}
\includegraphics[width=0.37\textwidth]{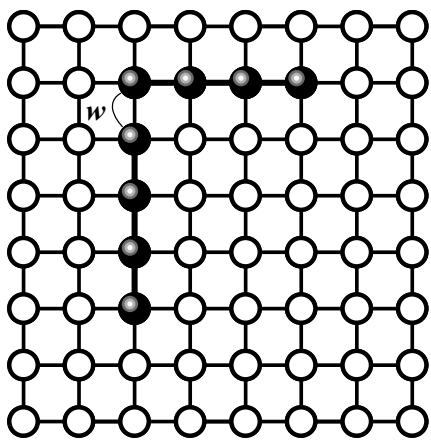}
\end{center}
\caption{Linear tetramers adsorbed on square lattices. Full and
empty circles represent tetramer units and empty sites,
respectively. } \label{figure1}
\end{figure}

We address the general case of adsorbates assumed to be linear
rigid particles containing $k$ identical units ($k$-mers), with
each one occupying a lattice site. Small adsorbates would
correspond to the monomer limit ($k = 1$). The distance between
$k$-mer units is assumed to be equal to the lattice constant;
hence exactly $k$ sites are occupied by a $k$-mer when adsorbed
(see Fig. \ref{figure1}). The surface is represented as an array
of $M = L \times L$ adsorptive sites in a square lattice
arrangement, where $L$ denotes the linear size of the array. In
order to describe the system of $N$ $k$-mers adsorbed on $M$ sites
at a given temperature $T$, let us introduce the occupation
variable $c_i$ which can take the values $c_i=0$ if the
corresponding site is empty and $c_i=1$ if the site is occupied.
On the other hand, molecules adsorb or desorb as one unit,
neglecting any possible dissociation. Under these considerations,
the Hamiltonian of the system is given by
\begin{equation}
H = w \sum_{\langle i,j \rangle} c_i c_j - N(k-1) w+ \epsilon_o
\sum_{i} c_i \label{h}
\end{equation}
where $w$ is the nearest-neighbor (NN) interaction constant which
is assumed to be attractive (negative), $\langle i,j \rangle$
represents pairs of NN sites and $\epsilon_o$ is the energy of
adsorption of one given surface site. The term $N(k-1)w$ is
subtracted in eq. (\ref{h})  since the summation over all the
pairs of NN sites overestimates the total energy by including
$N(k-1)$ bonds belonging to the $N$ adsorbed $k$-mers. Because the
surface was assumed to be homogeneous, the interaction energy
between the adsorbed dimer and the atoms of the substrate
$\epsilon_o$ was neglected for the sake of simplicity.

In order to characterize the phase transition, we use the order
parameter defined in Ref.~\cite{EPL1}, which in this case can be
written as
\begin{equation}
\delta =  \frac{\left | {n}_h - {n}_v \right |}{ {n}_h
 +  {n}_v }
 \label{fi}
\end{equation}
where $n_h(n_v)$ is the number of rods aligned along the
horizontal (vertical) direction. When the system is disordered
$(\theta<\theta_c)$, all orientations are equivalents and $\delta$
is zero. As the density is increased above $\theta_c$, the
$k$-mers align along one direction and $\delta$ is different from
zero. Thus, $\delta$ appears as a proper order parameter to
elucidate the phase transition.

The problem has been studied by grand canonical MC simulations
using a typical adsorption-desorption algorithm. The procedure is
as follows. Once the value of the chemical potential $\mu$ is set,
a linear $k$-uple of nearest-neighbor sites is chosen at random
and an attempt is made to change its occupancy state with
probability $W={\rm min} \left\{1,\exp\left(-\Delta H/k_BT \right)
\right\}$, where $\Delta H=H_f-H_i$ is the difference between the
Hamiltonians of the final and initial states and $k_B$ is the
Boltzmann constant. In addition, displacement (diffusional
relaxation) of adparticles to nearest-neighbor positions, by
either jumps along the $k$-mer axis or reptation by rotation
around the $k$-mer end, must be allowed in order to reach
equilibrium in a reasonable time. A MC step (MCs) is achieved when
$M$ $k$-uples of sites have been tested to change its occupancy
state. Typically, the equilibrium state can be well reproduced
after discarding the first $r'=10^7$ MCs. Then, the next $r=2
\times 10^7$ MCs are used to compute averages.

\begin{figure}
\begin{center}
\includegraphics[width=0.4\textwidth]{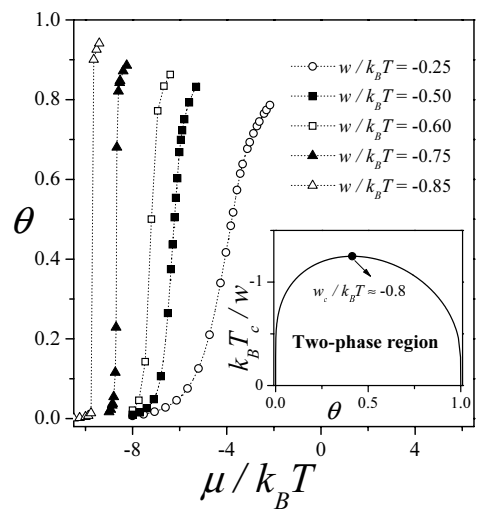}
\end{center}
\caption{Adsorption isotherms (coverage versus chemical potential)
for $k=10$, $L = 100$ different $w/k_BT$'s as indicated. Inset:
Adsorption phase diagram of attractive $10$-mers on square
lattices.} \label{figure2}
\end{figure}

In our MC simulations, we varied the chemical potential $\mu$ and
monitored the density $\theta$ and the order parameter $\delta$,
which can be calculated as simple averages. The reduced
fourth-order cumulant $U_L$ introduced by Binder~\cite{BINDER} was
calculated as:
\begin{equation}
U_L = 1 -\frac{\langle \delta^4\rangle} {3\langle
\delta^2\rangle^2}, \label{ul}
\end{equation}
where $\langle \cdots \rangle$ means the average over the MC
simulation runs. All calculations were carried out using the BACO
parallel cluster (composed by 60 PCs each with a 3.0 GHz Pentium-4
processor and 90 PCs each with a 2.4 GHz Core 2 Quad processor)
located at Instituto de F\'{\i}sica Aplicada, Universidad Nacional
de San Luis-CONICET, San Luis, Argentina.

\section{Results}

The calculations were developed for linear $10$-mers ($k=10$).
With this value of $k$ and for non-interacting rods, it is
expected the existence of a nematic phase at intermediate
densities~\cite{GHOSH}. The surface was represented as an array of
adsorptive sites in a square lattice arrangement with conventional
periodic boundary conditions. The effect of finite size was
investigated by examining lattices with $L=50, 100, 150, 200$.

In order to understand the basic phenomenology, we consider, in the
first place, the behavior of the adsorption isotherms in presence
of attractive lateral interactions between the $k$-mers.

Fig. \ref{figure2} shows typical adsorption isotherms (coverage
versus $\mu/k_BT$) for linear $10$-mers with different values of
the lateral interaction (the solid circles represent the Langmuir
case, $w/k_BT=0$).

The isotherms shift to lower values of chemical potential, and
their slopes increase as the ratio $w/k_BT$ increases (in absolute
value). For interaction strength above the critical value
($w>w_c$, in absolute values) the system undergoes a first-order
phase transition, which is observed in the clear discontinuity in
the adsorption isotherms\footnote{In this situation, which has
been observed experimentally in numerous systems, the only phase
which one expects is a lattice-gas phase at low coverage,
separated by a two-phase coexistence region from a
``lattice-fluid" phase at higher coverage. This condensation of a
two-dimensional gas to a two-dimensional liquid is similar to that
of a lattice-gas of attractive monomers. However, the symmetry
particle-vacancy (valid for monoatomic particles) is broken for
$k$-mers and the isotherms are asymmetric with respect to $\theta
= 0.5$.}. In the case studied, this critical value is
approximately $w_c/k_BT \approx -0.80$ (or $k_BT_c/w \approx
-1.25$). The behavior of the adsorption isotherms also allows us 
to calculate the phase diagram of the adsorbed monolayer in
``temperature-coverage" coordinates. In fact, once obtained the
real value of the chemical potential (or critical chemical
potential $\mu_c$) in the two-phase region, the corresponding
critical densities can be easily calculated. By repeating this
procedure for different temperatures ranging between $0$ and
$T_c$, the coexistence curve can be built~\cite{HILL}. A typical
phase diagram, obtained in this case for attractive $10$-mers, is
shown in the inset of Fig. \ref{figure2}.

\begin{figure}
\begin{center}
\includegraphics[width=0.40\textwidth]{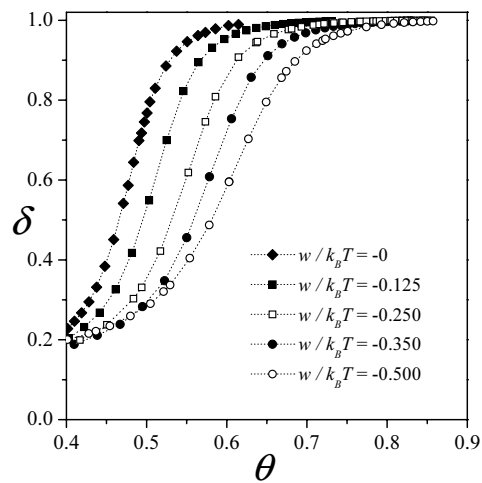}
\end{center}
\caption{Surface coverage dependence of the nematic order
parameter for $k=10$, $L = 100$ different $w/k_BT$'s as
indicated.} \label{figure3}
\end{figure}

On the basis of the study in Fig. \ref{figure2}, our next
objective is to obtain evidence for the existence of nematic order
in the range $-0.80 \leq w/k_BT < 0$ of attractive interactions.
For this purpose, the behavior of the order parameter $\delta$ as
a function of coverage was analyzed for $k=10$, $L=100$ and
different values of the lateral interaction. The results are shown
in Fig. \ref{figure3}, revealing that $(i)$ the orientational
order survives in the presence of attractive lateral interactions
and $(ii)$ the critical density shifts to higher values as the
magnitude of the lateral interactions is increased.

\begin{figure}
\begin{center}
\includegraphics[width=0.4\textwidth]{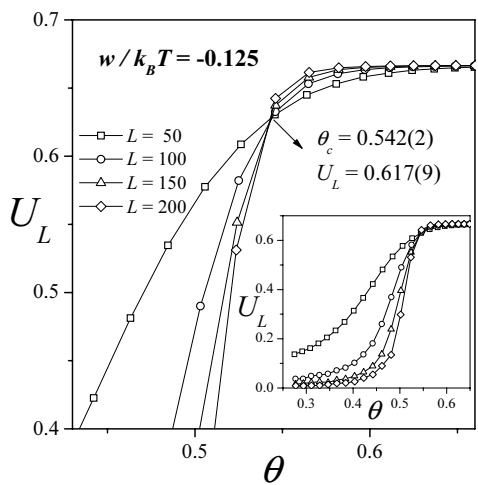}
\end{center}
\caption{Curves of $U_L(\theta)$ vs $\theta$ for $k=10$,
$w/k_BT=-0.125$ and square lattices of different sizes. From their
intersections one obtained $\theta_c$. In the inset, the data are
plotted over a wider range of densities.} \label{figure4}
\end{figure}

In order to corroborate the results obtained in the last figure,
we now study the dependence of $\theta_c$ on $w/k_BT$. In the case
of the standard theory of FSS~\cite{BINDER,PRIVMAN}, when the
phase transition is temperature driven, the technique allows for
various efficient routes to estimate $T_c$ from MC data. One of
these methods, which will be used in this case, is from the
temperature dependence of $U_L(T)$, which is independent of the
system size for $T=T_c$. In other words, $T_c$ is found from the
intersection of the curve $U_L(T)$ for different values of $L$,
since $U_L(T_c)=$const. In our study, we modified the conventional
FSS analysis by replacing temperature by density~\cite{EPL1}.
Under this condition, the critical density has been estimated from
the plots of the reduced four-order cumulants $U_L(\theta)$
plotted versus $\theta$ for several lattice sizes. As an example,
Fig. \ref{figure4} shows the results for $w/k_BT=-0.125$. In this
case, the value obtained was $\theta_c=0.542(2)$. In the inset,
the data are plotted over a wider range of temperatures,
exhibiting the typical behavior of the cumulants in the presence
of a continuous phase transition.

\begin{figure}
\begin{center}
\includegraphics[width=0.40\textwidth]{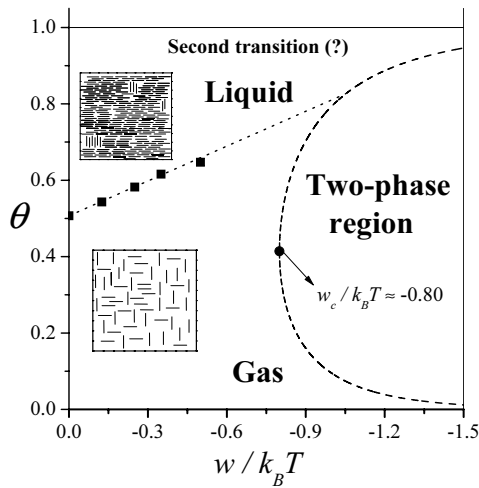}
\end{center}
\caption{Temperature-coverage phase diagram corresponding to
attractive $k$-mers with $k=10$. The inset in the upper-left
(lower-right) corner shows a typical configuration in the nematic
(isotropic) phase.} \label{figure5}
\end{figure}

The procedure of Fig. \ref{figure4} was repeated for $-0.80 \leq
w/k_BT < 0$, showing that the values of $\theta_c$ increase
linearly with the magnitude of the lateral couplings (see solid
squares in Fig. \ref{figure5}). The critical line (dotted line in
the figure) was obtained from the linear fit of the numerical
data. As it is possible to observe, the range of coverage at which
the transition occurs diminishes as $w/k_BT$ is increased (in
absolute value). This finding indicates that the presence of
attractive lateral interactions between the rods does not favor
the formation of nematic order in the adlayer. The phenomenon can
be understood from the behavior of the second virial coefficient,
which will initially decrease on introducing attractive $w$. This
decrease implies that the isotherms shift to lower values of
chemical potential, and consequently, the critical point shifts to
higher densities.

We did not assume any particular universality class for the
transitions analyzed here in order to calculate their critical
densities, since the analysis relied on the order parameter
cumulant's properties. However, the fixed value of the cumulants,
$U^* =0.617(9)$, is consistent with the extremely precise transfer
matrix calculation of $U^* =0.6106901(5)$~\cite{KAMIE} for the 2D
Ising model. This finding may be taken as an indication that the
phase transition belongs to the 2D Ising universality class.

\begin{figure}
\begin{center}
\includegraphics[width=0.35\textwidth]{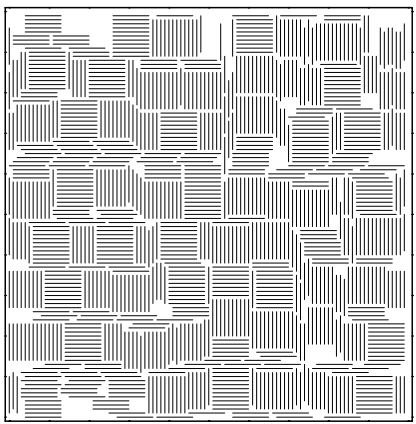}
\end{center}
\caption{Typical configuration of the adlayer in the high-coverage
phase and $w/k_BT < -0.80$.} \label{figure6}
\end{figure}

With respect to the behavior of the system for $w/k_BT < -0.80$,
the adsorbed layer ``jumps" from a low-coverage phase to a
high-coverage phase. This effect, which has been discussed in Fig.
\ref{figure2}, is represented in Fig. \ref{figure5} by the dashed
coexistence line. The low-coverage phase is an isotropic state,
similar to that observed for $w/k_BT
> -0.80$ and low density (see inset in the lower-right corner of Fig.
\ref{figure5}). On the other hand, the high-coverage phase is also
an isotropic state, but characterized by the presence of local
orientational order (domains of parallel $k$-mers). A typical
configuration in this regime is shown in Fig. \ref{figure6}).

Finally, it is worth pointing out that: $(1)$ the behavior of the
order parameter in Fig. \ref{figure3} clearly indicates that the
transition from the low-density disordered phase to the
intermediate-density ordered phase is an isotropic to nematic
phase transition (when all the words have the usual meaning). In
this case, the transition under study belongs to the 2D Ising
universality class. It can also be thought of as an unmixing or
liquid-gas transition \cite{FISCHER}. For this reason we have
called gas and liquid to the phases reported in Fig.
\ref{figure5}; and $(2)$ even though it has not been rigorously
proved yet, a second phase transition for non-interacting rods at
high densities has been theoretically predicted~\cite{GHOSH} and
numerically confirmed~\cite{JSTAT1}. This result has not been
confirmed for the case of attractive rods. An exhaustive study on
this subject will be the object of future work.

\section{Conclusions}

We have addressed the critical properties of attractive rigid rods
on square lattices with two allowed orientations, and shown the
dependence of the critical density on the magnitude of the lateral
interactions $w/k_BT$. The results were obtained by using MC
simulations and FSS theory.

Several conclusions can be drawn from the present work. On the one
hand, we found that even though the presence of attractive lateral
interactions between the rods does not favor the formation of
nematic order in the adlayer, the orientational order survives in
a range that goes from $w/k_BT=0$ up to $w_c/k_BT \approx -0.80$
($w_c/k_BT$ represents the critical value at which occurs a
typical transition of condensation in the adlayer). In this region
of $w/k_BT$, the critical density increases linearly with the
magnitude of the lateral couplings. On the other hand, the
evaluation of the fixed point value of the cumulants
$U^*=0.617(9)$ indicates that, as in the case of non-interacting
rods, the observed phase transition belongs to the universality
class of the two-dimensional Ising model.

With respect to the behavior of the system for $w/k_BT < -0.80$,
the continuous transition becomes first order. Thus, the adsorbed
layer jumps from a low-coverage phase, similar to that observed
for $w/k_BT > -0.80$ and low density, to an isotropic phase at
high coverage, characterized by the presence of local
orientational order (domains of parallel $k$-mers)

Future efforts will be directed to $(1)$ extend the study to
repulsive lateral interaction between the $k$-mers; $(2)$ obtain
the whole phase diagram in the space (temperature-coverage-rod's
size); $(3)$ develop an exhaustive study on critical exponents and
universality and $(4)$ characterize the second phase transition
from a nematic order to a non-nematic state occurring at high
density.

\begin{acknowledgements}
This work was supported in part by CONICET (Argentina) under
project number PIP 112-200801-01332; Universidad Nacional de San
Luis (Argentina) under project 322000 and the National Agency of
Scientific and Technological Promotion (Argentina) under project
33328 PICT 2005.
\end{acknowledgements}

\end{document}